\documentclass{moriond}

\def\Journal#1#2#3#4{{#1} {\bf #2}, #3 (#4)}

\def\PLB{{\em Phys. Lett.}  B}
\def\PRL{\em Phys. Rev. Lett.}

\def\be{\begin{equation}}
\def\ee{\end{equation}}
\def\bea{\begin{eqnarray}}
\def\eea{\end{eqnarray}}

\newcommand{\Photo}{}

\begin{document}
\vspace*{4cm}
\title{STANDARD MODEL SOFT QCD AT ATLAS AND CMS}

\author{ D. CAFORIO, on behalf of the ATLAS and CMS Collaborations}

\address{II. Physikalisches Institut, Justus-Liebig-Universit\"at, Heinrich-Buff-Ring 16,\\
35392 Giessen, Germany}

\maketitle\abstracts{
Recent results in soft QCD at LHC by the ATLAS, CMS and TOTEM Collaborations are presented.
Special focus is reserved to studies in diffractive and forward physics, and to the underlying event,
with comparison with previous results and highlighting novel techniques.}

\section{Diffractive and forward physics}
\subsection{Hard color-singlet exchange in dijet event}
In 2$\rightarrow$2 parton scattering, in the high-energy limit of QCD
the fixed-order perturbation theory approach is no longer valid.
In dijet production, the expected dynamics described by the Balitsky-Fadin-Kuraev-Lipatov (BFKL) evolution equation is reached in configurations where the two jets are separated by a large rapidity interval.

Events in proton-proton (pp) collisions with two jets separated by a large pseudorapidity ($\eta$) interval devoid of particle activity, known as Mueller-Tang jets or jet-gap-jet events, have been recently studied~\cite{hcse} by the CMS Collaboration~\cite{cms}.
The pseudorapidity gap is indicative of an underlying \textit{t}-channel hard color-singlet exchange.
In the BFKL framework, hard color-singlet exchange is described by \textit{t}-channel two-gluon ladder exchange between the interacting partons,
where the color charge carried by the exchanged gluons cancel, leading to a suppression of particle production between the final-state jets.

Soft rescattering effects between partons and the proton remnants can induce the production of particles in the $\eta$ interval that would otherwise be devoid of particles, resulting in a reduction of the number of events identified as having a jet-gap-jet signature.
Soft rescattering effects can be suppressed in processes where one or both of the colliding protons remain intact after the interaction, such as in single- or central-diffractive dijet processes or in dijet photoproduction, and can be used to better separate events with a central gap between the jets.
Jet-gap-jet events with an intact proton were also studied, using the TOTEM detector~\cite{totem}.

Occasionally charged particles are emitted at large angles with respect to the jet boundary into the $\eta$ region that should be devoid of particles: color-singlet exchange events appear as an excess of events over the expected charged particle multiplicity contribution from color-exchange dijet events at the lowest charged particle multiplicity.
The fraction of color-singlet exchange dijet events $f_{CSE}=\frac{N^F-N^F_{non-CSE}}{N}$ is defined as the difference between the number of dijets events ($N^F$) and the number of dijet events with no underlying color-singlet exchange ($N^F_{non-CSE}$), both at lowest multiplicity, over the total number of dijet events ($N$).

\begin{figure}
\begin{minipage}{0.5\linewidth}
\centerline{\includegraphics[width=0.9\linewidth]{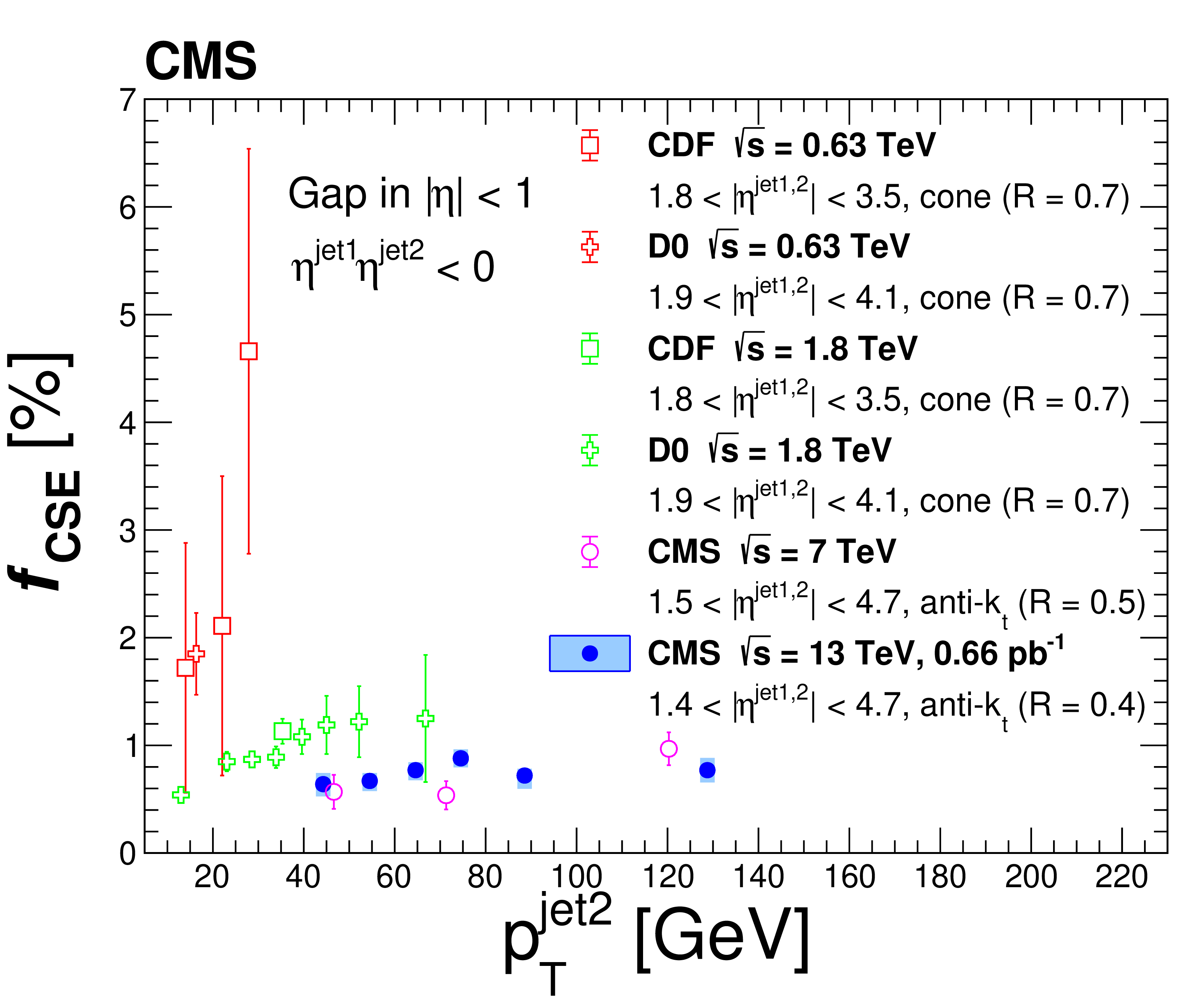}}
\end{minipage}
\hfill
\begin{minipage}{0.5\linewidth}
\centerline{\includegraphics[width=0.9\linewidth]{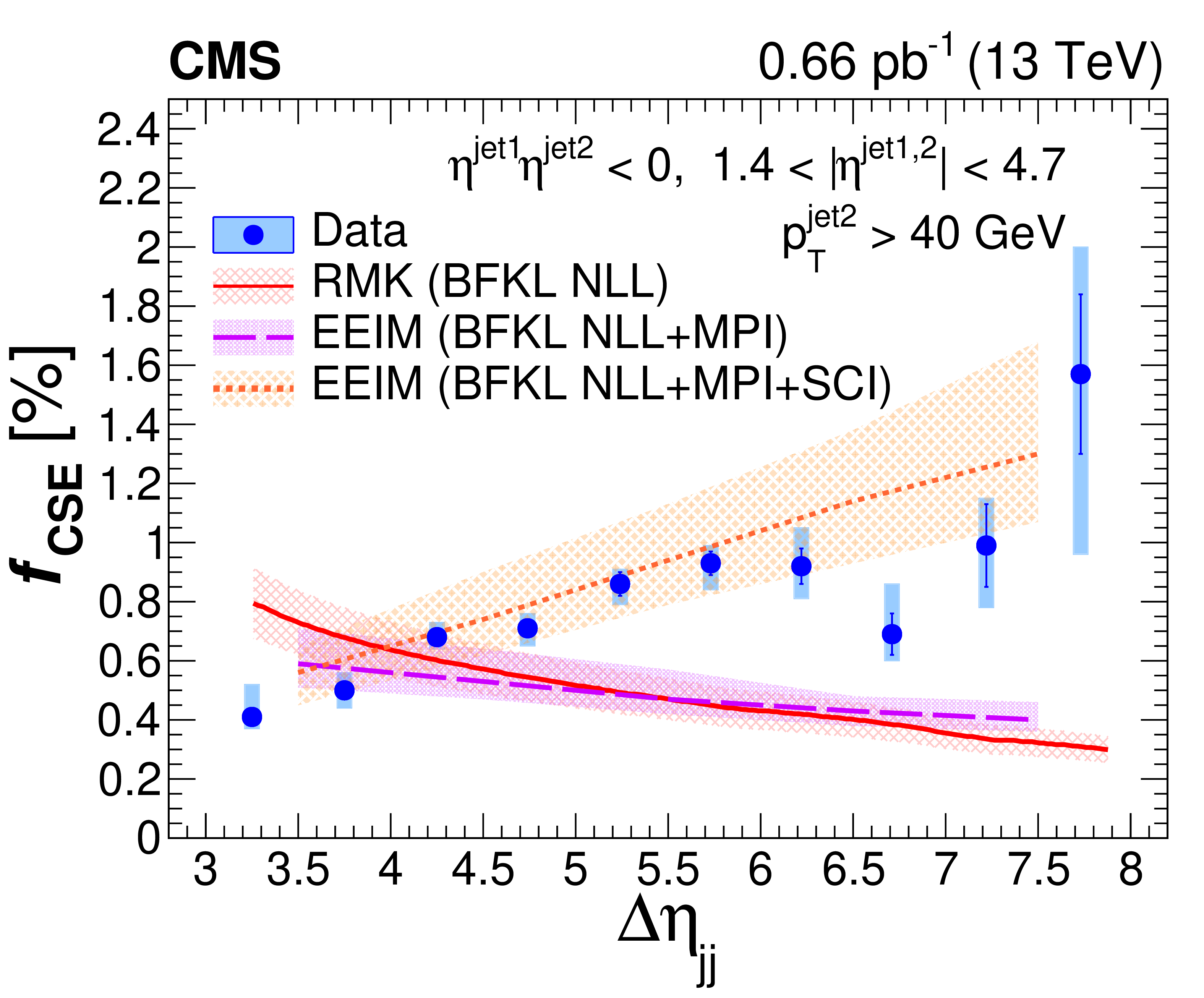}}
\end{minipage}
\caption[]{(Left) Fraction of color-singlet exchange dijet events, $f_{CSE}$, measured as a function of $p_T^{jet2}$ by the D0 and CDF Collaborations at $\sqrt{s}$ = 0.63 (red open symbols) and 1.8 TeV (green open symbols), by the CMS Collaboration at 7 TeV (magenta open symbols), and the present results at 13 TeV (filled circles).
  (Right) Fraction of color-singlet exchange dijet events, $f_{CSE}$, measured as a function of $\Delta\eta_{jj}$ in pp collisions at $\sqrt{s}$ = 13 TeV.
  The red solid curve corresponds to theoretical predictions based on the RMK model with gap survival probability of $\mid S \mid^2<$ = 10\%.
  The EEIM model predictions with MPI-only contributions and $\mid S \mid^2<$ = 1.2\% or MPI+SCI are represented by the purple dashed and orange dotted curves, respectively.
  The bands around the curves represent the associated theoretical uncertainties.~\cite{hcse}}
\label{fig:CSE_CMS3}
\end{figure}

The left plot on Figure~\ref{fig:CSE_CMS3} shows that there is no significant difference between 13 TeV and 7 TeV results, in contrast to the trend found at lower energies, an indication that the rapidity gap survival probability stops decreasing at the center-of-mass energies probed at the LHC.
The right plot on Figure~\ref{fig:CSE_CMS3} shows that the implementation by Ekstedt, Enberg, Ingelman and Motyka gives a fair description of the data within the uncertainties only when considering survival probability effects based on MPI and their soft color interaction model.
This is the first time that jet-gap-jet events were studied in association with an intact proton: the fraction of color-singlet exchange dijet events is almost three times larger than that for inclusive dijet production in dijets with similar kinematics, and this can be interpreted in terms of a lower spectator parton activity in events with intact protons, which decreases the likelihood of the central gap signature.

\subsection{Observation of forward proton scattering in association with lepton pairs produced via photon fusion}
Electromagnetic fields sourced by protons at the Large Hadron Collider (LHC) are sufficiently intense to exceed the Schwinger limit of $10^{18}$ V m$^{-1}$ and produce lepton pairs via photon fusion.
The ATLAS Collaboration~\cite{atlas} recently released a paper~\cite{phot} reporting the observation of forward proton scattering in association with lepton pairs produced via photon fusion.
The scattered protons are detected by the ATLAS Forward Proton spectrometer (AFP~\cite{afp1}~\cite{afp2}), whereas the leptons are reconstructed by the central ATLAS detector.
The dominant sources of background, estimated using a data-driven method, come from the Drell-Yan mechanism and by protons that are outside the AFP acceptance or were not reconstructed in AFP.

The measured fiducial cross sections in the $ee$ and $\mu\mu$ channels are $\sigma^{fid}_{ee+p}=11.0 \pm 2.6$ (stat) $\pm 1.2 $ (syst) $\pm0.3$ (lumi) fb and $\sigma^{fid}_{\mu\mu+p}=7.2 \pm 1.6$ (stat) $\pm 0.9$ (syst) $\pm 0.2$ (lumi) fb, respectively.
Table~\ref{tab:1} shows a comparison between these results and the proton soft survival models, where these factors are poorly constrained, especially at high $\gamma\gamma$ invariant masses that are important for new physics searches, because existing probes indirectly deduce dissociation rates using only central-detector information.
\begin{table}[t]
\caption[]{Fiducial cross sections from the combined HERWIG and LPAIR predictions with $S_{surv} = 1$ and $S_{surv}$ estimated using Refs.~\cite{harland}~\cite{dyndal}.
  SUPERCHIC 4 predictions include fully kinematically dependent $S_{surv}$.~\cite{phot}}
\label{tab:1}
\vspace{0.4cm}
\begin{center}
\begin{tabular}{ l c r }
\hline
\hline
  $\sigma_{HERWIG+LPAIR}\times S_{surv}$ & $\sigma^{fid}_{ee+p}$ (fb) & $\sigma^{fid}_{\mu\mu+p}$ (fb) \\
  \hline
  $S_{surv} = 1$              & 15.5$\pm$1.2 & 13.5$\pm$1.1 \\
  $S_{surv}$ using Refs.~\cite{harland}~\cite{dyndal} & 10.9$\pm$0.8 & 9.4$\pm$0.7 \\
  SUPERCHIC 4            & 12.2$\pm$0.9 & 10.4$\pm$0.7 \\
  Measurement                & 11.0$\pm$2.9 & 7.2$\pm$1.8 \\
\hline
\end{tabular}
\end{center}
\end{table}

\subsection{Measurement of differential cross sections for single diffractive dissociation in $\sqrt{s}$ = 8 TeV pp collisions using the ATLAS ALFA spectrometer}
Cross sections related to diffractive dissociation have been measured using early LHC data by exploiting the large rapidity gap signature that is kinematically expected.
These measurements are not able to distinguish fully between the single diffractive process, its double dissociation analogue in which both protons dissociate, and the tail of non-diffractive contributions in which large rapidity gaps occur.
The ATLAS Collaboration recently released a measurement of the single diffractive cross-section at 8 TeV ~\cite{xsec} with the ALFA forward proton spectrometer~\cite{alfa}.
In this process, the intact proton is scattered through a very small angle of typically a few tens of $\mu$rad, and is measured in the ALFA forward spectrometer, whereas the dissociated proton is measured in the tracking detector.
The interest in this measurement is that the double diffractive and non diffractive contributions are suppressed and it allows to measure the cross-section differentially in terms of the squared four-momentum transfer $t$, the fractional energy loss of the intact proton $\xi$ and the visible rapidity gap $\Delta\eta$.
Background arises mostly from central diffraction events.
As shown in Table~\ref{tab:2}, the generators describe the shape but overestimate the cross section, especially \textsc{Herwig7}.
\begin{table}[t]
\caption[]{The SD cross section within the fiducial region (4.0 $< \log_{10}\xi\le$ 1.6 and 0.016 $< \mid t \mid\le$
0.43 GeV$^2$) and extrapolated across all $t$ using the measured slope parameter $B$. The systematic
and statistical uncertainties are combined for data. The MC statistical uncertainties are negligible.~\cite{xsec}}
\label{tab:2}
\vspace{0.4cm}
\begin{center}
\begin{tabular}{ c  c  c }
  \hline
  \hline
  Distribution & $\sigma^{fiducial(\xi, t)}_{SD}$ [mb] & $\sigma^{t-extrap}_{SD}$ [mb] \\
\hline
  Data                                      & 1.59$\pm$0.13 & 1.88$\pm$0.15 \\
  \textsc{Pythia8} A2 (Schuler-Sj\"ostrand) & 3.69 & 4.35 \\
  \textsc{Pythia8} A3 (Donnachie-Landshoff) & 2.52 & 2.98 \\
  \textsc{Herwig7}                          & 4.96 & 6.11 \\
\hline
\end{tabular}
\end{center}
\end{table}

\section{The underlying event}
\subsection{Underlying Event in top pairs}
Another typical soft QCD phenomenon is the underlying event.
The underlying event model in $t\bar{t}$ events has been tested up to a scale of two times the top quark mass in a study performed by the CMS Collaboration~\cite{uetop}, and measurements in categories of dilepton invariant mass indicate that it should be valid at even higher scales.
These measurements are also relevant as a direct probe of color reconnection, which is needed to confine the initial QCD color charge of the $t$ quark into color-neutral states.
The main contribution to the underlying event comes from the color exchanges between the beam particles and is modeled in terms of multiparton interactions, color reconnection and beam-beam remnants, tuned to minimum bias and Drell-Yan data.
The main background arises from $tW$ and Drell-Yan processes.
In $t\bar{t}$ events the contribution of the underlying event is typically order of 20 charged particles with transverse momentum of about 2 GeV.
The majority of the distributions analyzed indicate a fair agreement between the data and \textsc{PowegPythia8}, and disfavor the setups in which multiparton interactions and color reconnection are switched off, or in which the strong coupling parameter is increased.
They also disfavor default configurations in \textsc{PowegHerwig} and \textsc{Sherpa}.
The value of the strong coupling parameter at the mass of $Z$ boson $\alpha^{FSR}_S(M_Z)=0.120\pm0.006$ is consistent with the data, and the corresponding uncertainties translate to a variation of the renormalization scale by a factor of $\sqrt{2}$.

\subsection{Underlying Event in $Z\rightarrow\mu^+\mu^-$}
ATLAS recently released a paper~\cite{uez} studying the underlying event in events containing a $Z$ boson decaying into a muon pair.
Charged-particle multiplicities and transverse momentum are measured in regions of the azimuth defined relative to the $Z$ boson direction.
The absence of QCD final-state radiation allows the study of different kinematic regions with varying transverse momenta of the $Z$ boson.
All tested generators show significant deviations from the data, but predict the mean values better when focusing on the MPI-sensitive regions.

\vspace*{1cm}
Copyright 2021 CERN for the benefit of the ATLAS and CMS Collaborations. CC-BY-4.0 license.

\end{document}